\documentstyle[seceq,epsbox,preprint]{jpsj}

\def\runtitle
{Memory Effect, Rejuvenation and Chaos Effect in the MREM}
\def\runauthor{Munetaka {\sc Sasaki} and Koji {\sc Nemoto}}
\title{Memory Effect, Rejuvenation and Chaos Effect in the Multi-layer Random Energy Model}
\author{Munetaka {\sc Sasaki} and Koji {\sc Nemoto}}

\inst{Division of Physics, Hokkaido University, Sapporo 060-0810}
\recdate{\phantom{October 4, 1999}}
\abst{
We introduce magnetization to the Multi-layer Random Energy Model
which has a hierarchical structure, and perform Monte Carlo 
simulation to observe the behavior of ac-susceptibility.
We find that this model is able to reproduce three prominent features
of spin glasses, i.e., memory effect, rejuvenation and chaos effect,
which were found recently
by various experiments on aging phenomena with temperature variations.
}
\kword{aging, multi-layer random energy model, hierarchical picture, memory effect, rejuvenation, chaos effect }
\begin{document}
\sloppy
\maketitle
\section{Introduction}\label{sec:introduction}
The free energy of complex systems such as spin glasses,
 polymers and proteins, is considered
 to have a very complex structure with numerous local minima. 
Due to this complexity,
 it costs very long time for equilibration of these systems and
various non-equilibrium relaxation phenomena are observed. 
From experiments focused on these non-equilibrium relaxations, 
the aging phenomena, which are dynamical behaviors
 largely depending on the history of system, were found and 
have been studied vigorously from both theoretical 
and experimental aspects, in particular, in spin glass systems
\cite{ZFCexperiment1,ZFCexperiment2,TRMexperiment,ZFCTS,ZFCTC,MCS}.

Recently the following interesting experiment has been reported 
in spin glasses\cite{acTC1,acTC2,acTC3}. 
In this experiment, the relaxation of the out-of-phase 
ac-susceptibility \( \chi'' \) is observed in the following 
three stage. In the first stage, the sample is quenched from above 
\( T_{\rm c} \) down to a temperature \( T \) below \( T_{\rm c} \) 
and is kept at this temperature during \( t_1\). 
Then the sample is perturbed by changing temperature from \( T \) 
to \( T\pm \Delta T\) during \( t_2\) in the following second
stage and the temperature is returned to \( T \) (the third stage). 
The effect of the perturbation is examined by comparing the 
perturbed data and unperturbed (\( t_2 =0 \)) data 
in the third stage. In the case \( -\Delta T \), both data coincide 
except at the very beginning. This means that the relaxation 
at \( T -\Delta T \) never affects the relaxation at \( T \) and 
the system remembers the relaxation during \( t_1 \) after 
perturbation. We call this as memory effect hereafter. 
In the case \( +\Delta T \), on the other hand, 
we can see that \( \chi'' \) becomes larger than the unperturbed 
correspondence and the rejuvenation occurred during \( t_2 \). 
These 'memory effect' and 'rejuvenation' are observed in other 
experiments\cite{TRMexperiment}. 

From the theoretical point of view,
 the aging phenomena have been studied along 
two different pictures. One is the droplet 
picture\cite{droplet1,droplet2}, where the behavior in the 
spin glass phase is governed by the low laying and large length scale 
excitation. The other is the so-called hierarchical picture
\cite{hierarchical1,hierarchical2,hierarchical3,hierarchical4,Bouchaud}, 
from which  we will try to understand the phenomena in this manuscript.

According to the Parisi mean-field solution of the SK model, 
the free energy in the spin-glass phase has a very complex structure 
with numerous local minima. These valleys are considered to 
be divided into smaller valleys as the system is cooled 
and micro phase transitions occur with this continuous multifurcation. 
From this hierarchical structure,
 the memory effect and rejuvenation mentioned above can be
explained as follows\cite{acTC2}. 
When the temperature is cooled from temperature above \( T_{\rm c} \) 
down to \( T \), the system falls into one of the numerous local minima. 
Then, the system is equilibrated in some region of the phase space 
including several valleys, \( {\cal R} \), during \( t_1 \). 
Now we consider to change the temperature from \( T \) to 
\( T\pm\Delta T \). In the case \( -\Delta T \), 
each valley at \( T \) is divided into smaller valleys. 
If \( \Delta T\) is large enough and the time \( t_2 \) 
is not so long, the equilibration proceeds only in the smaller valleys 
and we can ignore the possibility that the system overcomes 
the barrier which separates the bigger valleys at \( T \). 
Therefore, when the temperature is raised back to 
\( T \) and smaller valleys merge, we find that 
the system are unchanged during \( t_2 \) and 
the memory effect appears. In the case \( +\Delta T \), on the other hand, 
several valleys in \( {\cal R} \) merge into one valley 
and the system forges the equilibration at \( T \). 
Thus the system is rejuvenated. 

This explanation is, however, quite qualitative
 and there exists no theoretical model which reproduces
 these features along this hierarchical picture.
 In this manuscript, we study the 
Multi-layer Random Energy Model (MREM)\cite{Bouchaud,MREM2,MREM3} 
which  has a hierarchical structure causing a continuous phase transition. 
We introduce  magnetization to this model so that we can carry out 
the simulation to observe the ac-susceptibility, and
find  that the memory effect and rejuvenation are
 indeed reproducible. 
Furthermore, we analyze the mechanism of these phenomena and 
show that the above-mentioned scenario realizes these features, 
at least, in this model. 

Another feature we are interested in is the {\em chaos} effect found
in experiments on the ac-susceptibility\cite{Jonason}.
The fact found is that
the equilibration at the temperature \( T \) does not affect 
the relaxation at a  lower temperature \( T-\Delta T \).
This can be interpreted as follows.
If the structure of the free energy at \( T-\Delta T \)
 is quite different from that at \( T \), the equilibration 
 at \( T \) never helps the equilibration at \( T-\Delta T \). 
But this reorganization of structure makes us unable to explain
 the memory effect, which is also observed in the same experiment.
There seem to exist few theoretical 
explanations
for these apparently conflicting 
features, i.e., memory and {\em chaos} effects.
We find that these conflicting features hold 
simultaneously in the MREM in a  situation slightly different 
from the experiment in ref.~\citen{Jonason}. 
We also analyze the mechanism of these phenomena in this model. 

\section{The Multi-layer Random Energy Model}
Let us first explain the Simple-layer Random Energy Model (SREM)
illustrated in Fig.~1.
 The bottom points represent the accessible states of this system. 
We hereafter consider the case that the number of states is very large. 
The length of each branch represents the barrier energy $E$, over which
the system goes from one state to another.
This energy is an independent random variable distributed as
\begin{equation}
\rho (E) {\rm d}E = \frac{{\rm d} E}{T_{\rm c}}\exp[-E/T_{\rm c} ],
\label{eqn:simplerho}
\end{equation}
where \( T_{\rm c} \) is the transition temperature. 

From the Arrhenius law, 
the relaxation time \( \tau(\alpha) \), i.e., the average time
 for the system to escape from the state \( \alpha \), 
is related to \( E(\alpha) \) as
\[ \tau(\alpha) = \tau_0 \exp[ E(\alpha)/T], \]
where \( \tau_0 \) is a microscopic time scale. From 
eq.~(\ref{eqn:simplerho}), the distribution of \( \tau \) can be written as
\begin{equation}
p(\tau) {\rm d} \tau = \frac {x \tau_0^x}{\tau^{x+1}}{\rm d} \tau 
\hspace{1cm}(\tau \geq \tau_0),
\label{eqn:sinpleptau}
\end{equation}
where \( x \equiv T/T_{\rm c} \).
From eq.~(\ref{eqn:sinpleptau}), 
it is easily shown that the averaged relaxation time \( \langle\tau\rangle \) 
is \( \tau_0 x/(x-1) \) for \( x>1 \) and infinite for \( x \le 1 \). 
This means that the transition from the ergodic phase to the non-ergodic phase 
occurs at \( T_{\rm c} \). 

Now we construct the Multi-layer Random Energy Model (MREM). As shown in 
Fig.~2,
 this model is constituted by piling up the SREM $L$ times hierarchically. 
The energy of the \( n \)-th layer counted from the bottom, \( E_n \),
 is given according to the 
distribution \( \rho_n(E_n)=\exp[-E_n/ T_{\rm c}(n)]/T_{\rm c}(n) \) 
and each layer has a different transition temperature chosen so as to satisfy
\( T_{\rm c}(1)<T_{\rm c}(2)<\cdots <T_{\rm c}(L) \). 
Therefore, in this model the transition occurs continuously 
from the uppermost (the \( L \)-th) layer to the lowest one.

\subsection{The dynamics}
In this simulation, we use the following Markoff process for dynamics. 
First the system is thermally activated from \( \alpha \) to 
\( \alpha_1 \), where \( \alpha_n \) is the $n$-th {\em ancestor}
 of \( \alpha \) (see Fig.~2).
The probability that 
this event occurs at time $t$ in unit time is defined as
\begin{equation}
\omega(\alpha;t) = \tau_0^{-1} \exp[-E_1(\alpha)/T].
\label{eqn:omega}
\end{equation}
The $t$ dependence  comes implicitly from time variation of
temperature and  magnetic field. 
We will introduce the effect of magnetic field in the following subsection. 
Since the time evolution is Markoffian, we can easily evaluate
the probability $q(\alpha;t_0,t){\rm d}t$ for
the event to occur during
\( t \) and \( t+{\rm d} t \)
 with knowing that the system is in the state $\alpha$ at time $t_0$ as
\begin{equation}
q(\alpha;t_0,t){\rm d}t=\omega(\alpha;t){\rm d}t\exp[-\int_{t_0}^{t}{\rm d}t'
\omega(\alpha;t')].
\label{eqn:Etime}
\end{equation}
We generate an event time $t$ according to \( q(\alpha;t_0,t) \) to
perform event-driven Monte Carlo simulation. 

Once the event occurs and the system is activated to $\alpha_1$,
it may be activated further to 
\( \alpha_2 \), $\alpha_3$ or higher branching point. 
We accept the excitation to \( \alpha_2 \) with the probability 
\( \exp[-E_2(\alpha)/T] \).
If it is accepted, the next trial to $\alpha_3$ follows.
By repeating these procedures, we obtain
the probability \( H(n) \) that the system is activated from $\alpha_1$
up to $\alpha_n$ as
\begin{equation}
H(n) =
 \left(1-\exp[-E_{n+1}(\alpha)/T]\right) \prod_{k=2}^n \exp[-E_k(\alpha)/T],
\label{eqn:eqforHL}
\end{equation}
where \( E_{L+1}(\alpha) \equiv \infty \). 

Finally, we adopt the simplest falling process from \( \alpha_n \). 
The system falls to one of all states under \( \alpha_n \) with 
equal probability. In the limit that the number of branches in 
each layer is infinite, we can neglect the possibility that the system has 
visited one state more than twice in our finite time simulation. 
Therefore, in this simulation, we create a new 
state \( \beta \) after each activation. Note that if the system 
is activated to $n$-th layer, 
we set \( E_k(\beta)=E_k(\alpha) \) for \( k \ge n+1 \), although 
for \( k\le n \), \( E_k \) is updated according to the density of state 
\( \rho_k(E) \).

We note that the whole process drives the system to the equilibrium
distribution proportional to $\exp[\sum_n E_n/T]$.

\subsection{Assignment of magnetization}
In order to observe the magnetic response, we introduce magnetization to the MREM.
It is natural to suppose that
the nearer two states are located in the phase space
  the stronger the correlation of the two magnetizations is, and that 
   the distance in the MREM can be measured in terms of barrier height.
In fact the barrier height is related with the overlap in the SK model\cite{Nemoto}.
To incorporate this aspect, we assign the value of magnetization to state $\alpha$, $M_\alpha$, as
\begin{equation}
M_{\alpha} ={\cal M}_0(\alpha)+{\cal M}_1(\alpha_1)+\cdots
+{\cal M}_{L-1}(\alpha_{L-1}),
\end{equation}
where \( {\cal M}_k(\alpha_k) \) is a contribution 
from the branch point $\alpha_k$ given as a independent random variable with zero mean.
The correlation between $M_\alpha$ and $M_\beta$ with the lowest common branch point $\alpha_k$
comes from the common contributions of ${\cal M}_n, (n=k,k+1,\cdots)$ to these magnetization.
It decreases monotonically as $k$ increases and the barrier becomes higher.

In this simulation, we choose a uniform distribution for ${\cal M}$ of the $n$-th layer as
\begin{equation} 
D_n({\cal M}) = \left\{ 
  \begin{array}{cl}
\frac{\sqrt{L}}{2} &\mbox{($|{\cal M}| \leq \frac{1}{\sqrt{L}} $)}, \\
0 &\mbox{($ |{\cal M}| > \frac{1}{\sqrt{L}} $)}.
\end{array}\right.
\end{equation}
The range of distribution is determined so that 
the variance of $M_\alpha$ is independent of $L$.

The Zeeman energy due to applying magnetic field $H(t)$ is attributed to the lowest layer and
\( E_1(\alpha) \) in eq.~(\ref{eqn:omega}) is replaced
with \( E_1(\alpha)+H(t)M_{\alpha} \). 

To summarize, we introduce the concrete procedure of our simulation.
\begin{itemize}
\item[(i)] Choose the initial state \( \alpha \). Because we consider the 
situation that the system is quenched from infinitely high temperature, we 
determine \( E_k(\alpha) \) \((k=1,\ldots,L)\) according to \( \rho_k(E) \). 
We also determine \( {\cal M}_k(\alpha) \) from \( D_k({\cal M}) \). 
\item[(ii)]Determine the time \( t_{\rm stay} \) that the system stay at 
\( \alpha \) from eq.~(\ref{eqn:Etime}).
\item[(iii)]Determine how far the system is activated by eq.~(\ref{eqn:eqforHL}). 
\item[(iv)]Give the energy and magnetization of new state \( \beta \). 
When the system is activated to \( \alpha_n \), we set 
\( E_k(\beta)=E_k(\alpha) \) and \( {\cal M}(\beta)={\cal M}(\alpha) \) 
for \( k \ge n+1 \) and give \( E_k(\beta) \) and \( {\cal M}(\beta) \)
from \( \rho_k(E) \) and \( D({\cal M}) \) for \( k \le n \). 
\item[(v)] Set the time ahead by \( t_{\rm stay} \) and return to (ii).
\end{itemize}

We show an example of simulation for \( L=2 \) with a weak ac-field. 
We plot \( {\cal M}_k(t) \) \( (k=1,2) \) in one trial in Fig.3(a), and 
averaged magnetization over \( 2\times 10^7 \) trials in Fig.3(b). 
We can see that in one trial, \( {\cal M}_k \) keeps a constant during 
the system stays at a state and changes discontinuously when activation 
occurs. We can see that the activation to the \( 1 \)-st layer and the one 
to the \( 2 \)-nd layer occurs at \( t_2 \) and \( t_1 \) respectively. 
From this figure, we can not find the effect of ac-field. But this effect 
emerges by taking average over numerous trials, as shown in Fig.~3(b). 
We estimate \( \chi''(t) \) directly from data of this kind. 

\section{The Result of Simulations}
In the following we show the result of simulations performed on the MREM with
$L=2$, $T_{\rm c}(1)=0.6$ and $T_{\rm c}(2)=1.0$.
The number of samples used for random average is typically $10^7$.
The amplitude and the period of applying ac-field are fixed to $0.1$ and $100\tau_0$, respectively.
Hereafter this period is used as the unit of time. 

\subsection{The case $-\Delta T$}
In the first stage the system evolves in temperature $T=0.85$ for time interval $t_1$.
Temperature is then reduced to $T-\Delta T=0.5$ for $t_2$ (the second stage), 
and brought back to $T$ in the following third stage.
Note that these temperatures are set so as to satisfy
\[ T-\Delta T < T_{\rm c}(1) < T < T_{\rm c}(2). \]
In Fig.~4, we plot imaginary part of ac-susceptibility, $\chi''$,
 for two different intervals of the first stage, $t_1=20$ and $100$.
In inset we plot \( t_1\) and \( t_3\) part of data. 
We can see that the memory effect takes place. 
We also find  in the data for $t_1=100$ the slight jump
at the beginning of the second stage as seen in experiment\cite{acTC1}.

In Fig.~5, we plot \( \chi_0'' \) and \( \chi_1'' \)  evaluated 
from \( {\cal M}_0 \) and \( {\cal M}_1 \), respectively ($\chi''=\chi_0''+\chi_1''$). 
We can see that \( \chi_0'' \) remains almost constant in the 
first and third stage because  \( T>T_{\rm c}(1) \) 
so that the first layer quickly relaxes to the equilibrium.
The long-time dependence of relaxation is 
dominated by the second layer there.  In contrast the first layer is
dominant in the second stage because $T-\Delta T$ is too low for the second layer
to be activated and \( \chi_1'' \) is much smaller than 
\( \chi_0'' \).
 From these results, we notice that the dominant layer changes as
temperature varies. 

we can also see that the slight jump of \( \chi'' \) mentioned above 
is brought from \( \chi_0''\).
This sudden increase of 
\( \chi_0'' \) is due to the quenching of the first layer across its transition temperature \( T_{\rm c}(1) \). 

In Fig.~6, \( \chi_s'' \) in the second stage is magnified for $t_1=20$ and $t_1=100$. 
There are few differences on both \( \chi_0''\) and \( \chi_1'' \).
This means that the length of the first stage, \( t_1 \),
does not affect the relaxation in the second stage.
In this sense, we may say that the system has
the {\em chaos} effect.
On the other hand, \( t_1 \) reflects on relaxation 
in the third stage and the memory effect appears.
Thus the MREM has these two apparently conflicting 
features simultaneously. 

In order to investigate which states contribute to $\chi''$ in more detail,
we examine the energy distribution
 \( P_n(E_n,t) \) which is defined as the probability density
that the system is found at time $t$ in one of the states whose energy of the $n$-th layer is $E_n$.
By the definition, we have $P_n(E_n,t=0)=\rho_n(E_n)$ since the initial state is chosen randomly.

In Fig.~7, we show the time dependence of \( P_n(E_n,t) \) in the first stage.
We can see that the distribution consists of two exponential functions which are connected to each other 
at some point, $E^*$.
The value of $E^*$ roughly represents the energy level up to which the system can be activated during time
interval $t$, and is estimated as
\begin{equation}
E^*\approx T\log[t/\tau_0].
\end{equation}
For \( E_n \le E^* \) the distribution is aged or equilibrated, so that
the exponent $\alpha_1(n)$ is given as
\begin{equation}
\alpha_1(n)=\frac1T-\frac{1}{T_{\rm c}(n)},
\end{equation}
while the other part $E_n \ge E^*$ leaves untouched and
the exponent $\alpha_2(n)$ is equal to that of $\rho_n(E)$, i.e.,
\begin{equation}
\alpha_2(n)=-\frac{1}{T_{\rm c}(n)}.
\end{equation}

For the first layer we have $\alpha_1(1) < 0$ since \( T > T_{\rm c}(1) \), and
the distribution quickly converges to the equilibrium one (Fig.7, left).
Note that the discrepancy for $E_1>E^*$ is negligible in magnitude.
For the second layer where $\alpha_1(2)>0$, on the other hand, \( P_2(E_2,t) \) 
has a peak at $E^*$ moving right with time (Fig.~7, right).
This is why
\( \chi_1'' \) 
continues to change with time
while
  \( \chi_0'' \) remains almost constant
 in the first stage.

Experiments on the Zero-Field-Cooled (ZFC) magnetization in 
spin glasses\cite{ZFCexperiment1,ZFCexperiment2} showed that
the distribution of the relaxation time 
\( \tau \), which has a peak at $\tau_{\rm max}$,
 depends on the waiting time \( t_{\rm w} \) 
and  \( \tau_{\rm max} \) appears near \( t_{\rm w} \), which 
just corresponds to the shift of peak shown in the right of Fig.~7.

In Fig.~8, we show the time dependence of \(P_n(E_n,t)\) in the second stage. 
We can see  that a peak appears and moves to right for the first layer 
(Fig.~8, left).
As for \( P_2(E_2,t)\), the global aspects such as the position of
peak formed in the first stage do not change
although the distribution of the lower energy
decreases gradually. This brings the memory effect to the system. 

To see how $t_1$ affects the energy distribution, 
we plot in Fig.~9 $P_n(E_n,t)$ for 
\( t_1=20 \) and \( t_1=10^2 \) with the same $t$ in the second stage.
As for the first layer, the both distributions are 
almost the same since the layer is well equilibrated in the first stage.
In the second stage, this layer is dominant and 
the behaviors of \( \chi'' \) are almost the same in the two cases.
On the other hand, the difference of the peak position quenched since the end of first stage,
makes the shapes of
$P_2(E_2,t)$ very different from each other, and this causes, in turn, 
the difference in the behavior of $\chi''$ in the third stage.

\subsection{The case $+\Delta T$}
In Fig.~10, the $\chi''$ for the case \( +\Delta T \) is shown. The heating
is done twice in this case.
We set the temperatures as \( T=0.5\), \( T+\Delta T=0.85 \), which satisfy
the relation $T<T_{\rm c}(1)<T+\Delta T<T_{\rm c}(2)$.
We can see from this figure that the rejuvenation occurs.

We show the behavior of \( \chi_0'' \) and \( \chi_1'' \) in Fig.~11.
In this case 
we can see that the $\chi_1''$ dominates the slow relaxation in the second and fourth stages
as expected.
The rejuvenation occurs only in $\chi_0''$ but not in $\chi_1''$.

In Fig.~12, we compare the behavior of \( \chi'' \) in the first, third and 
fifth stages. Although the value in the first stage is slightly larger than that in the third,
those in the third and fifth stages are collapsed to a single curve,
which means the perfect re-initialization.

We again examine \( P_n(E_n,t) \) in this case.
 In Fig.~13, the time evolution of \( P_1(E_1,t) \) in the second stage is shown.
The peak created in the first stage is rapidly 
destroyed and the information on the relaxation 
 in the first stage is completely forgotten.
In fact we find that the peak starts to disappear at \( t\simeq t_1+1 \)
, i.e., just after the second stage begins. 

\section{Conclusions and Discussions}
We have shown that the present MREM can reproduce the prominent behaviors 
found in spin glasses such as the memory effect and the rejuvenation although
the number of layers is only two.
This indicates that the alternation of the activated layer is important to
explain these features in this model.
We expect that the phenomena can be observed at any $T<T_{\rm c}$
in the MREM with large $L$.

Now let us discuss what happens when $L\gg 1$ along the scenario proposed 
by Bouchaud and Dean\cite{Bouchaud}. 
For a given temperature $T<T_{\rm c}(L)=T_{\rm c}$, there exists the $n$-th layer
such that $T_{\rm c}(n-1)<T<T_{\rm c}(n)$.
The essential point is the fact that layers below $n$ are quickly equilibrated and
they forget what happened before the temperature is set, while those above $n$ are
almost quenched and they behave as if the time evolution stops.
In this sense the $n$-th layer is the {\em activated} one and dominates the relaxation
of the system.
This mechanism certainly explains the memory effect and the rejuvenation as shown in
the present work.
In this context, the {\em chaos} effect means that the layer which is to be activated when
temperature decreases is not capable to remember the previous situation and it relaxes from
{\em tabulae rasae}.
Therefore it is not necessary to introduce a {\em chaotic} reorganization of hierarchy
for description of the phenomena.


Finally, let us compare the droplet model with the MREM and try to 
fuse these models. In the droplet model, 
system {\em ages} by growth of droplets. On the other hand, aging means 
the shift of the peak of \( P(E) \) in the {\em frozen} layers 
(the layers which satisfy \( T < T_{\rm c}(n) \)) in the MREM. 
In both model, aging means to seek more stable states with time and 
these processes make the system stiffer and the response to 
an external field weaker. This is the reason why 
the ac-susceptibility decreases monotonously with age. 
Moreover, as discussed Jonason {\it et al}\cite{Jonason}, it may be possible 
to map the MREM into the droplet model as following. We now consider 
what will happen when we cool real spin glass systems across the 
\( T_{\rm c} \) (which corresponds to \(T_{\rm c}(L)\) in the MREM). Above 
\( T_{\rm c} \), all the spins flip rather free. When the system is cooled 
just below \( T_{\rm c} \), we can consider that larger droplets are blocked 
at first and the spins in these droplets begin to flip collectively. In this 
stage, the spins in smaller droplet are still free relatively. 
In the MREM, the larger droplets correspond to the upper {\em frozen} layers 
and the smaller droplets correspond to the lower {\em unfrozen} layers. 
As the temperature is lowered, smaller droplets begin to be blocked and flip 
collectively. But the larger droplets which blocked earlier are almost frozen 
and they can hardly flip at this temperature. 
We expect that these successive 
processes bring the memory effect, rejuvenation and {\em chaos} effect 
as discussed in this manuscript. Furthermore, the idea of 
{\em droplets within droplets}\cite{DPinDP} will naturally lead to the hierarchical 
organization of droplets. It is challenging to find these structures, 
if really exist, in the real spin space of spin glasses. 
\section*{Acknowledgements}
We are very grateful to H. Yoshino for fruitful discussions and
suggestions on the manuscript. The numerical calculations were performed 
on an Origin 2000 at Division of Physics, Graduate school of Science, 
Hokkaido University.

\newpage
\noindent
{\bf \large FIGURE CAPTIONS}

%
\vspace*{3mm}\noindent
Fig.~1  Structure of the Simple-layer Random Energy Model.

\vspace*{3mm}\noindent
Fig.~2  Structure of the Multi-layer Random Energy Model with $L=2$.

\vspace*{3mm}\noindent
Fig.~3  (a) ${\cal M}_1(t)$ (dashed line) and ${\cal M}_2(t)$ (solid line) 
        observed in one trial and (b) averaged 
        ${\cal M}_k(t)$ over $2\times 10^7$ trials. 

\vspace*{3mm}\noindent
Fig.~4  $\chi''$ for the case \( -\Delta T \) with
        \( t_1=20 \) (dashed line) and \( t_1=100 \) (solid line). 
        In the inset, the data in the first and third stage 
        are plotted.                                

\vspace*{3mm}\noindent
Fig.~5  \( \chi_0'' \) (solid line) and \( \chi_1'' \) (dashed line) for the 
        case $-\Delta T$.

\vspace*{3mm}\noindent
Fig.~6  $\chi''$ in the second stage. \( \chi_0'' \) with \( t_1=20 \), 
        \( \chi_0'' \) with \( t_1=100 \), \( \chi_1'' \) with  \( t_1=20 \)
        and \( \chi_1'' \) with \( t_1=100 \) (from top to bottom).

\vspace*{3mm}\noindent
Fig.~7  \( P_n(E_n,t) \) in the first stage at
        \( t=1,10^{0.5},10,\ldots,10^2 \) (from left to right).

\vspace*{3mm}\noindent
Fig.~8  \( P_n(E_n,t) \) in the second stage at
         \( t=0,10^{-1.5},10^{-1},\ldots,10^2 \) (from left to right).

\vspace*{3mm}\noindent
Fig.~9  \( P_n(E_n,t) \) for \( t_1=20 \) (left) and 
         \( t_1=100 \) (right) in the second stage at \( t=t_1+10^{1.5} \).

\vspace*{3mm}\noindent
Fig.~10  $\chi''$ for the case \( +\Delta T \). In the inset,
         the tree parts of perturbed data and unperturbed data are shown.

\vspace*{3mm}\noindent
Fig.~11  \( \chi_0'' \) (upper) and \( \chi_1'' \) (lower) for the case \( +
         \Delta T \).

\vspace*{3mm}\noindent
Fig.~12  \( \chi'' \) in the first, third and fifth stages.

\vspace*{3mm}\noindent
Fig.~13  \( P_1(E_1,t) \) in the second stage for the 
         case \( +\Delta T \) at time \( t=t_1+\Delta t \) with $\Delta t = 
         10^{-1.5},10^{-1},\ldots,10^{2}$.
\newpage
\begin{center}
\epsfile{file=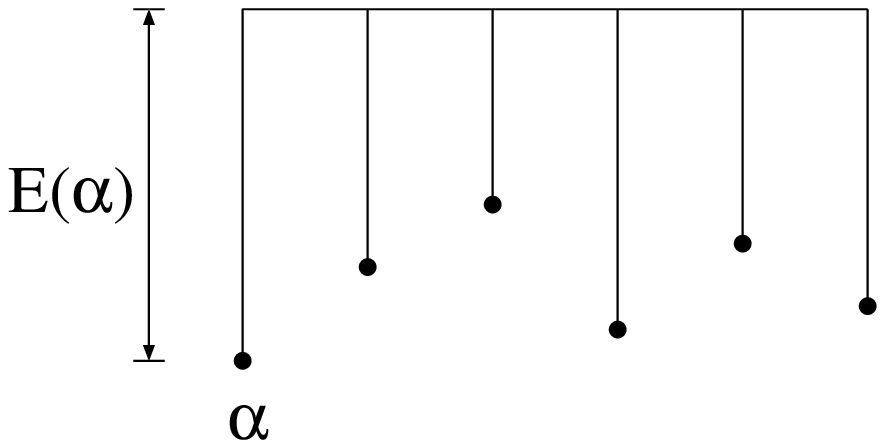,width=16.5cm}
\end{center}
\vspace{2cm}
\begin{center}
{\LARGE Fig.1}
\end{center}
\newpage
\begin{center}
\epsfile{file=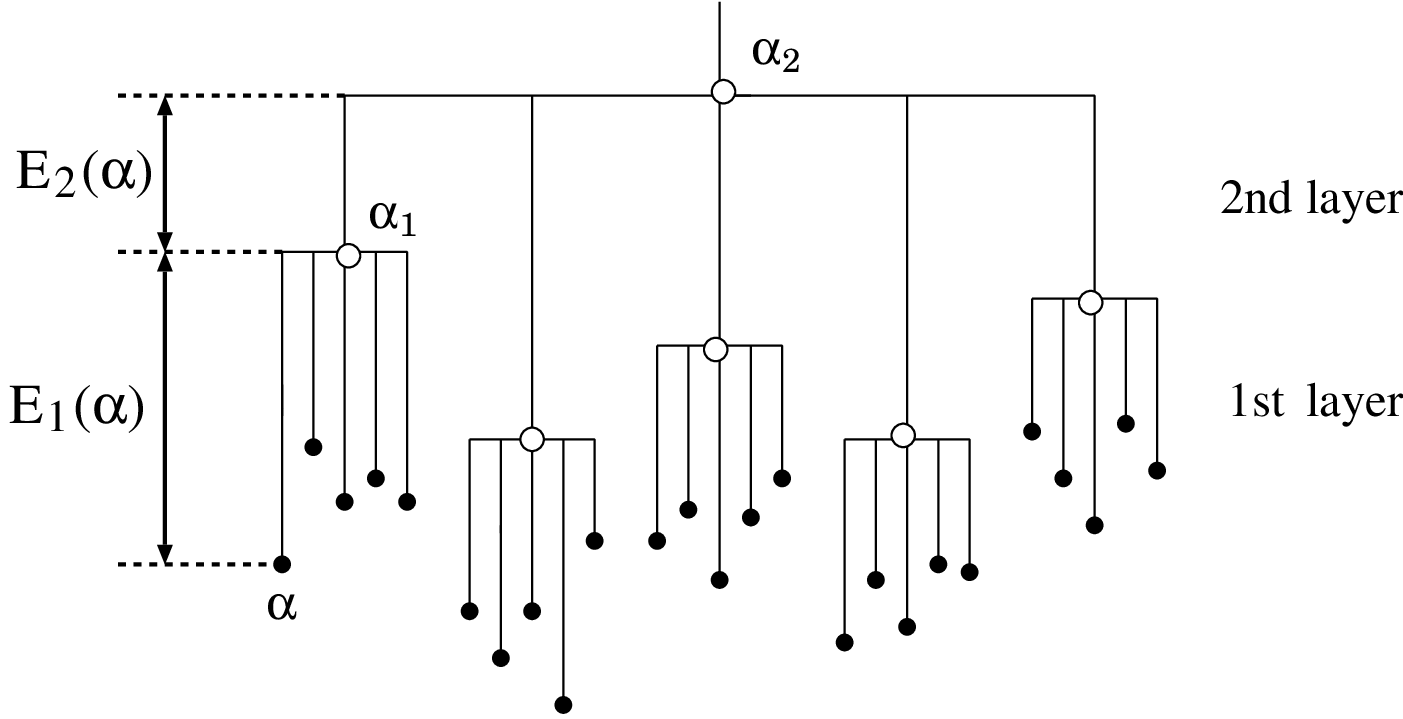,width=16cm}
\end{center}
\vspace{2cm}
\begin{center}
{\LARGE Fig.2}
\end{center}
\newpage
\begin{center}
\epsfile{file=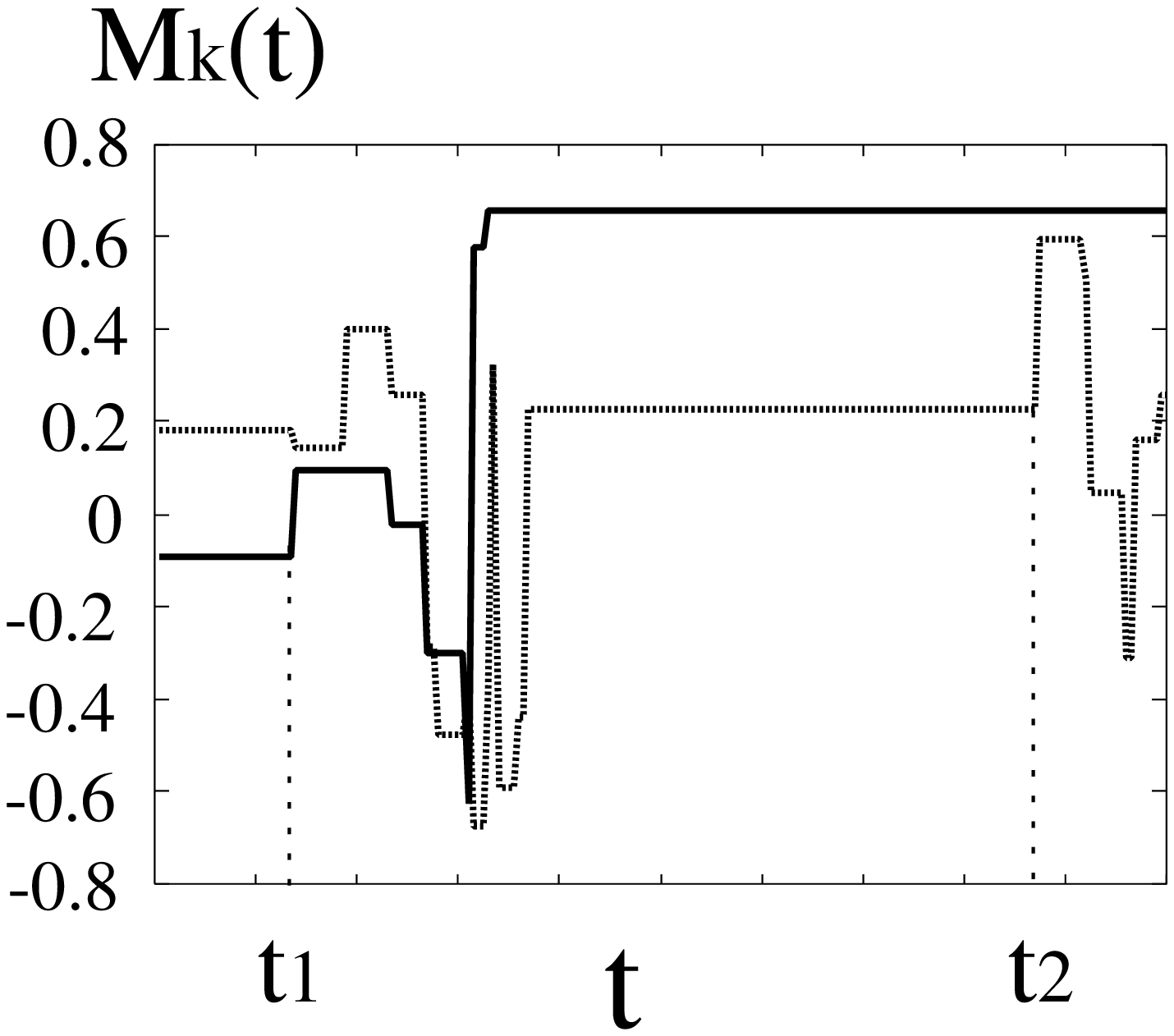,width=16.5cm}
\end{center}
\vspace{2cm}
\begin{center}
{\LARGE Fig.3(a)}
\end{center}
\newpage
\begin{center}
\epsfile{file=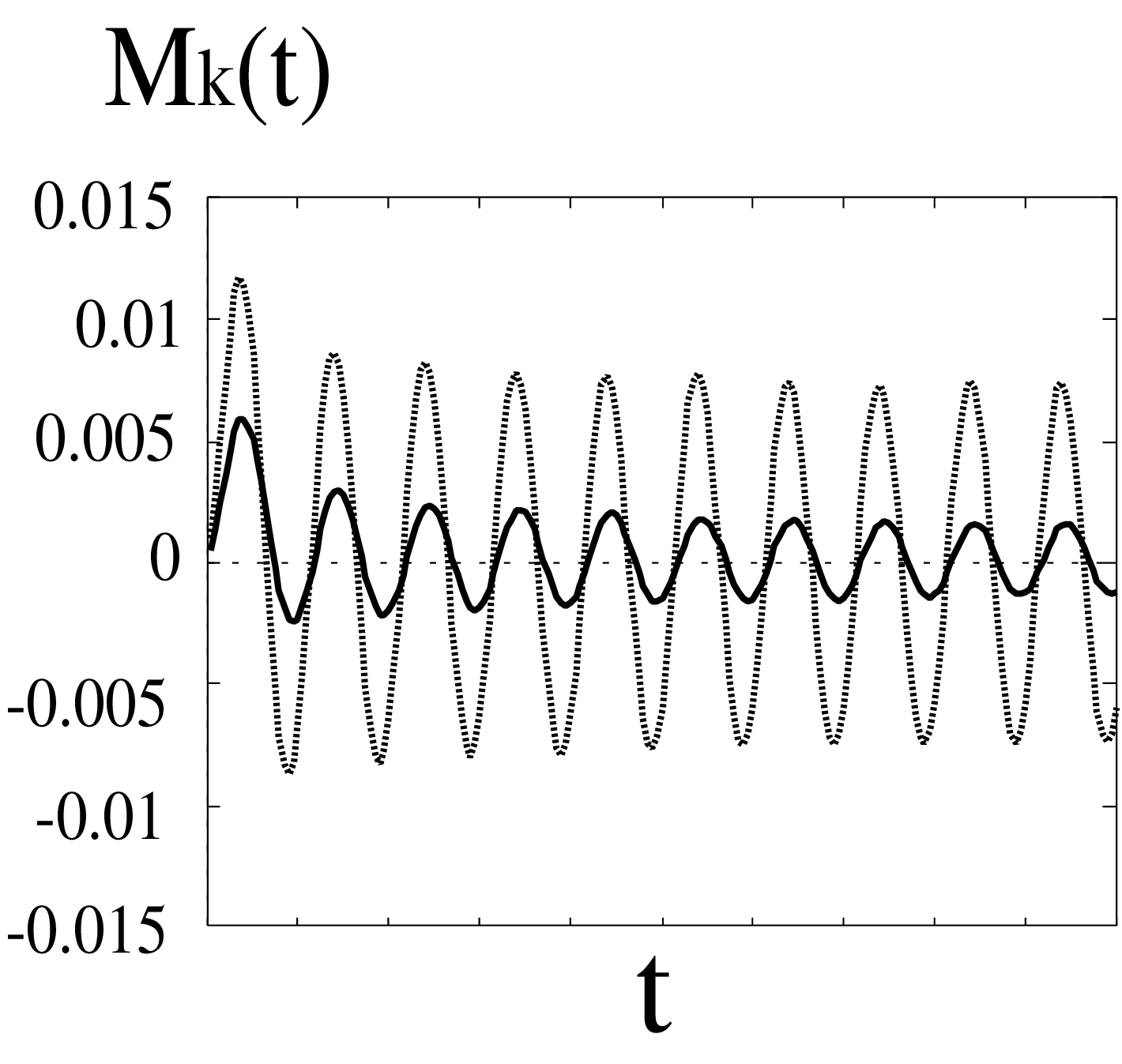,width=16.5cm}
\end{center}
\vspace{2cm}
\begin{center}
{\LARGE Fig.3(b)}
\end{center}
\newpage
\begin{center}
\epsfile{file=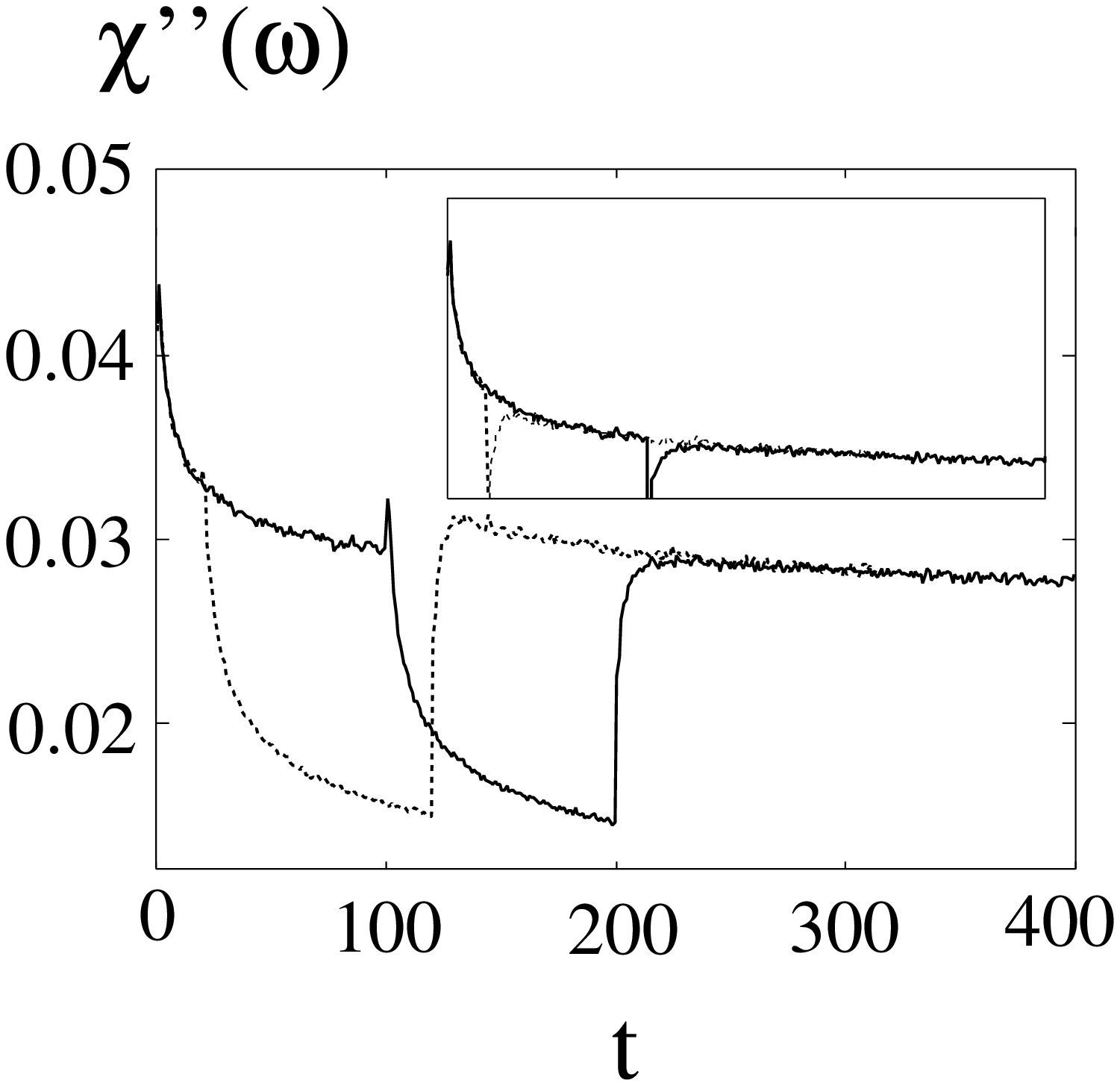,width=16.5cm}
\end{center}
\vspace{2cm}
\begin{center}
{\LARGE Fig.4}
\end{center}
\newpage
\begin{center}
\epsfile{file=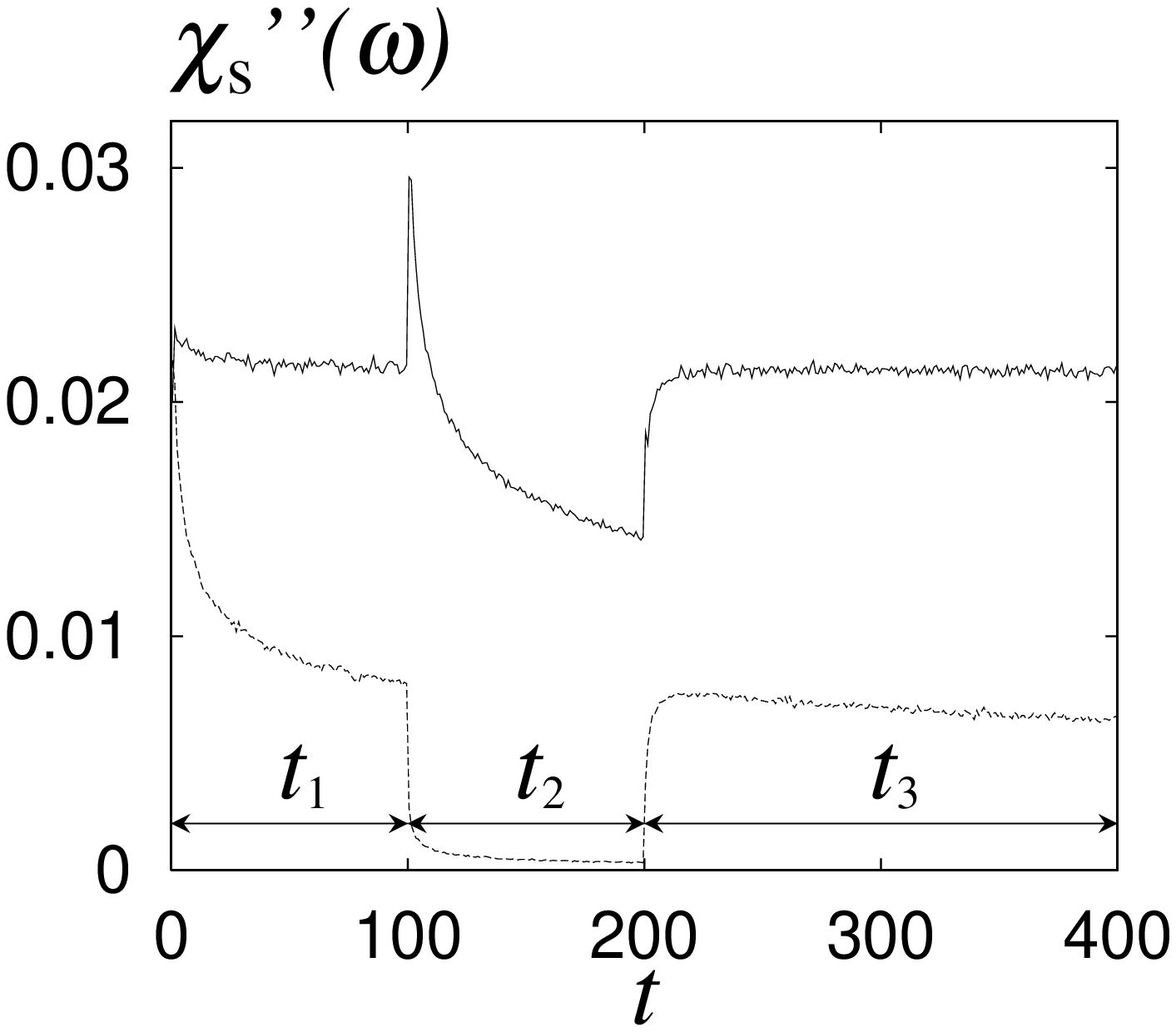,width=16.5cm}
\end{center}
\vspace{2cm}
\begin{center}
{\LARGE Fig.5}
\end{center}
\newpage
\begin{center}
\epsfile{file=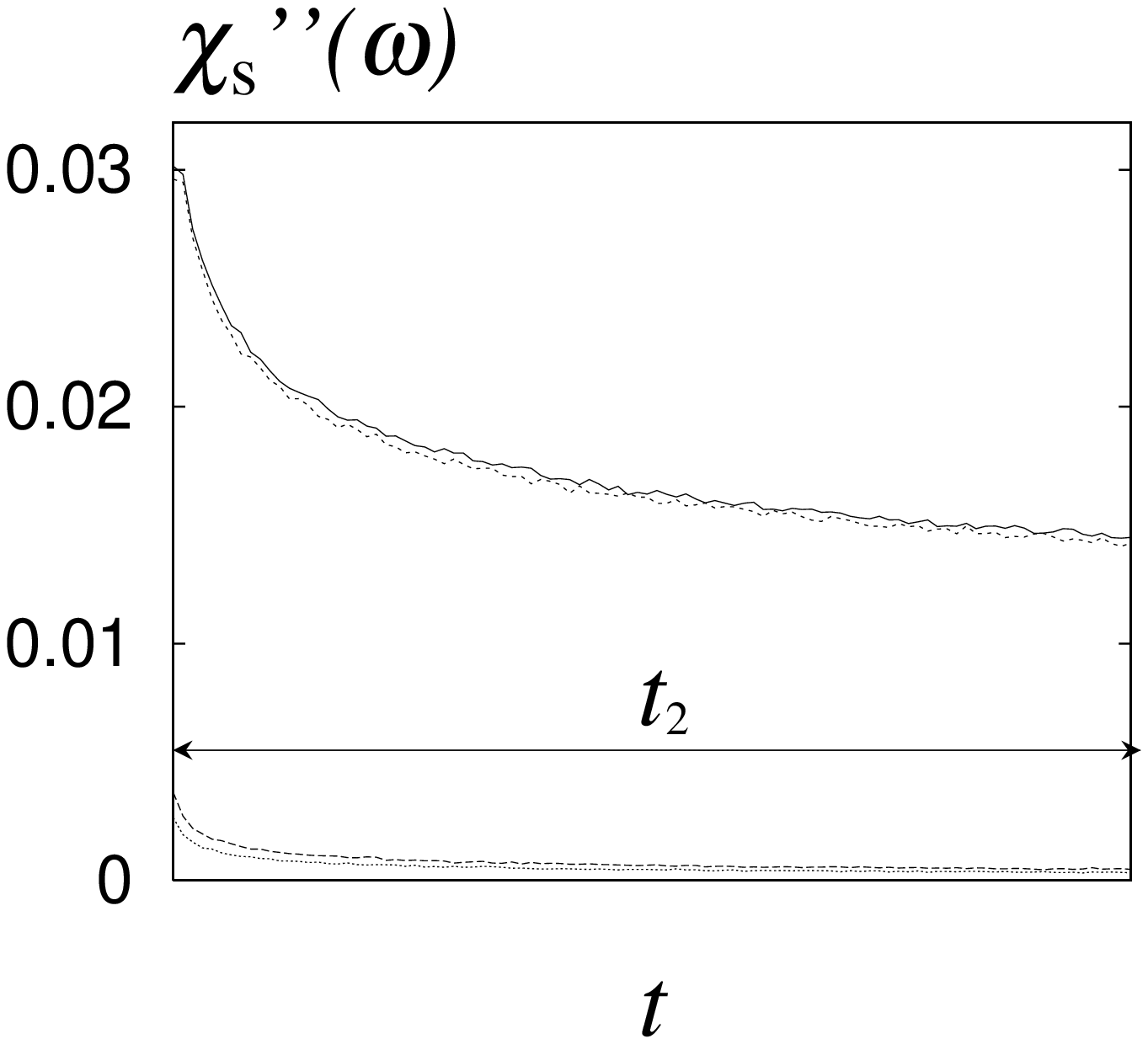,width=16.5cm}
\end{center}
\vspace{2cm}
\begin{center}
{\LARGE Fig.6}
\end{center}
\newpage
\begin{center}
\hspace*{-1cm}
\epsfile{file=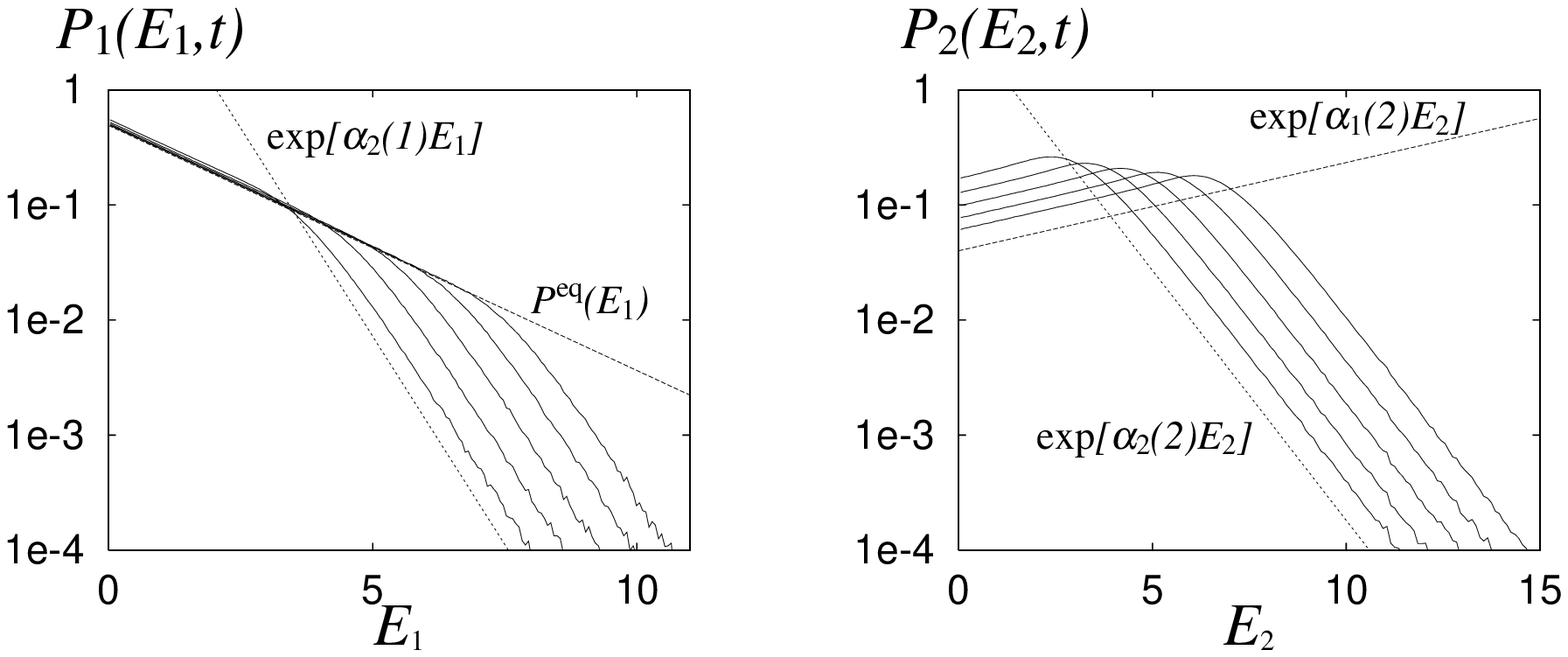,width=17.5cm}
\end{center}
\vspace{2cm}
\begin{center}
{\LARGE Fig.7}
\end{center}
\newpage
\begin{center}
\hspace*{-1cm}
\epsfile{file=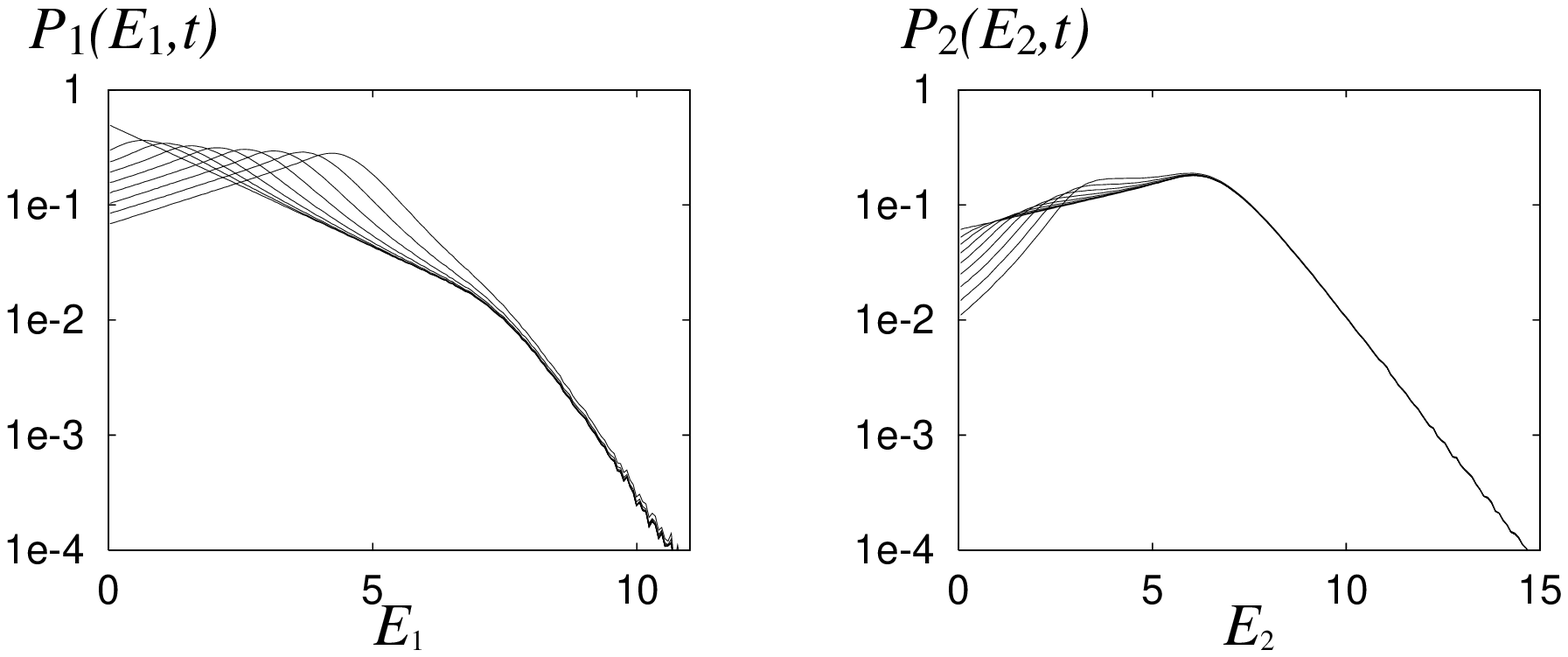,width=17.5cm}
\end{center}
\vspace{2cm}
\begin{center}
{\LARGE Fig.8}
\end{center}
\newpage
\begin{center}
\hspace*{-1cm}
\epsfile{file=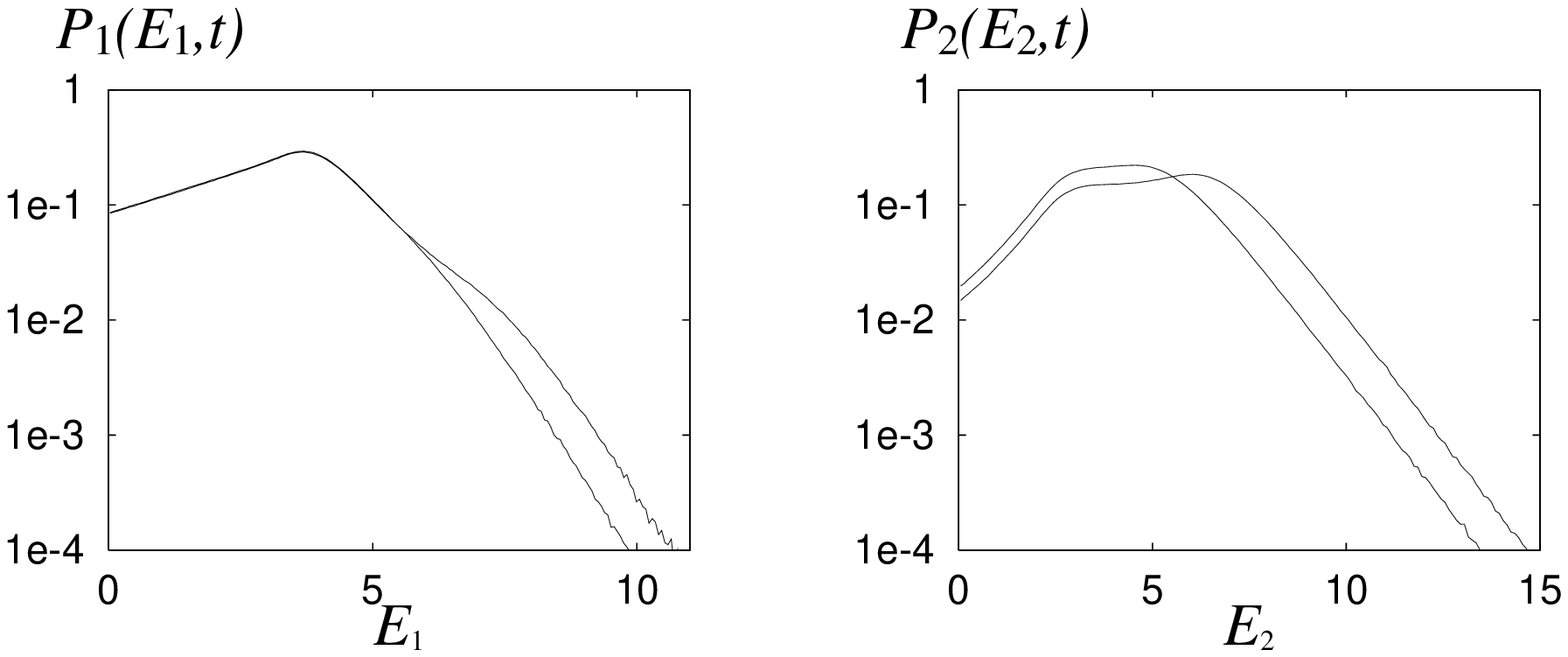,width=17.5cm}
\end{center}
\vspace{2cm}
\begin{center}
{\LARGE Fig.9}
\end{center}
\newpage
\begin{center}
\epsfile{file=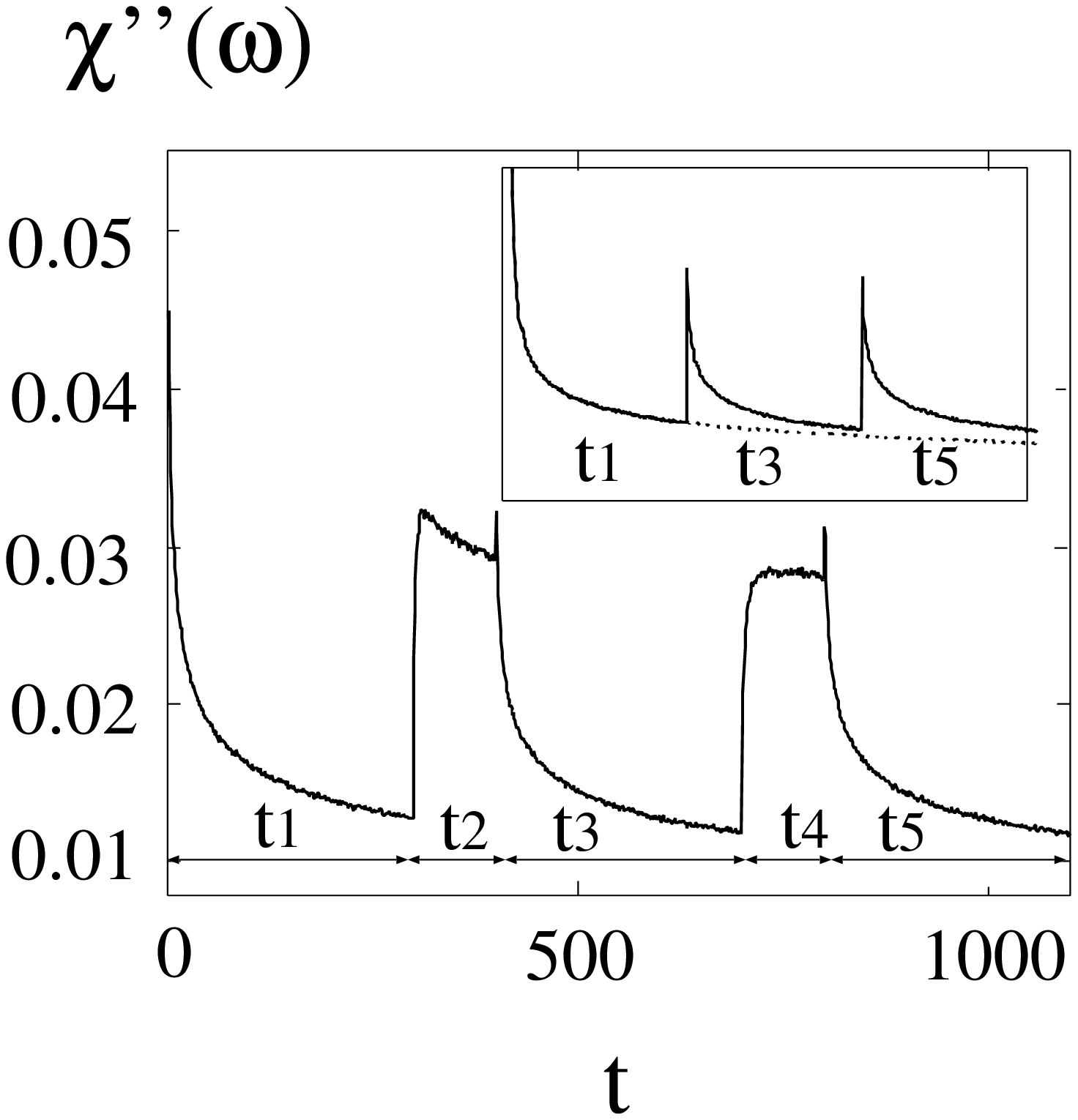,width=16.5cm}
\end{center}
\vspace{2cm}
\begin{center}
{\LARGE Fig.10}
\end{center}
\newpage
\begin{center}
\epsfile{file=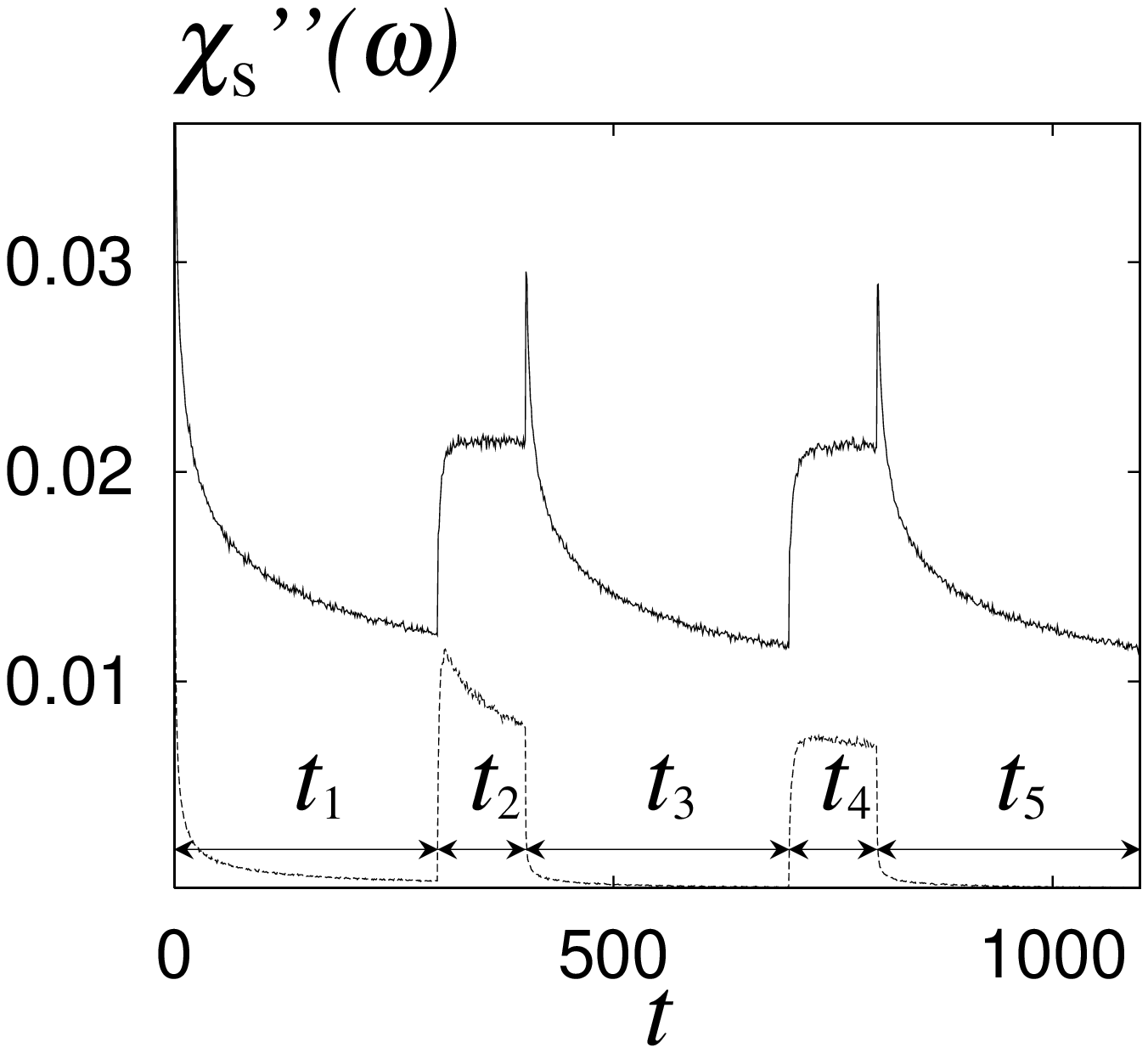,width=16.5cm}
\end{center}
\vspace{2cm}
\begin{center}
{\LARGE Fig.11}
\end{center}
\newpage
\begin{center}
\epsfile{file=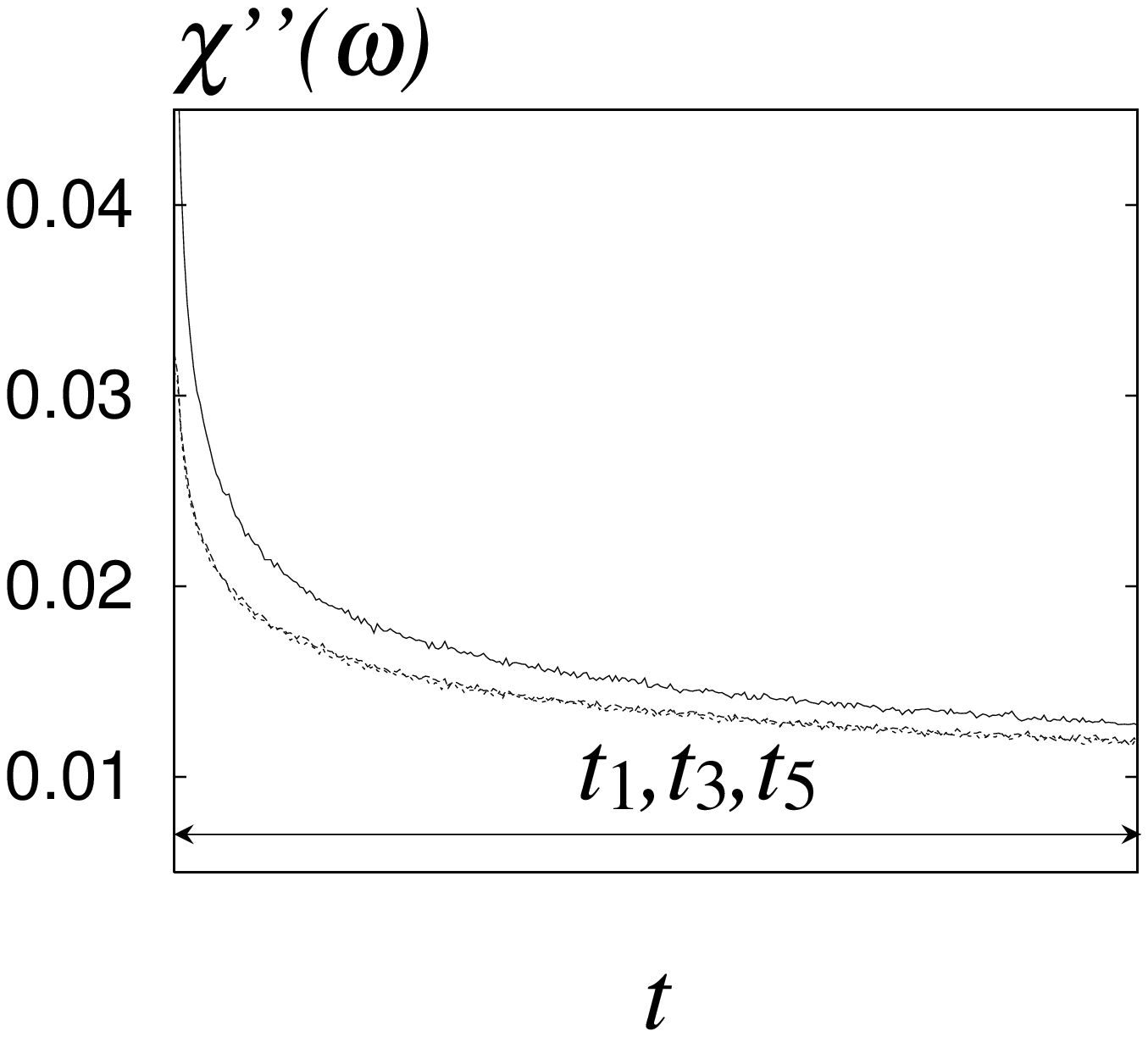,width=16.5cm}
\end{center}
\vspace{2cm}
\begin{center}
{\LARGE Fig.12}
\end{center}
\newpage
\begin{center}
\epsfile{file=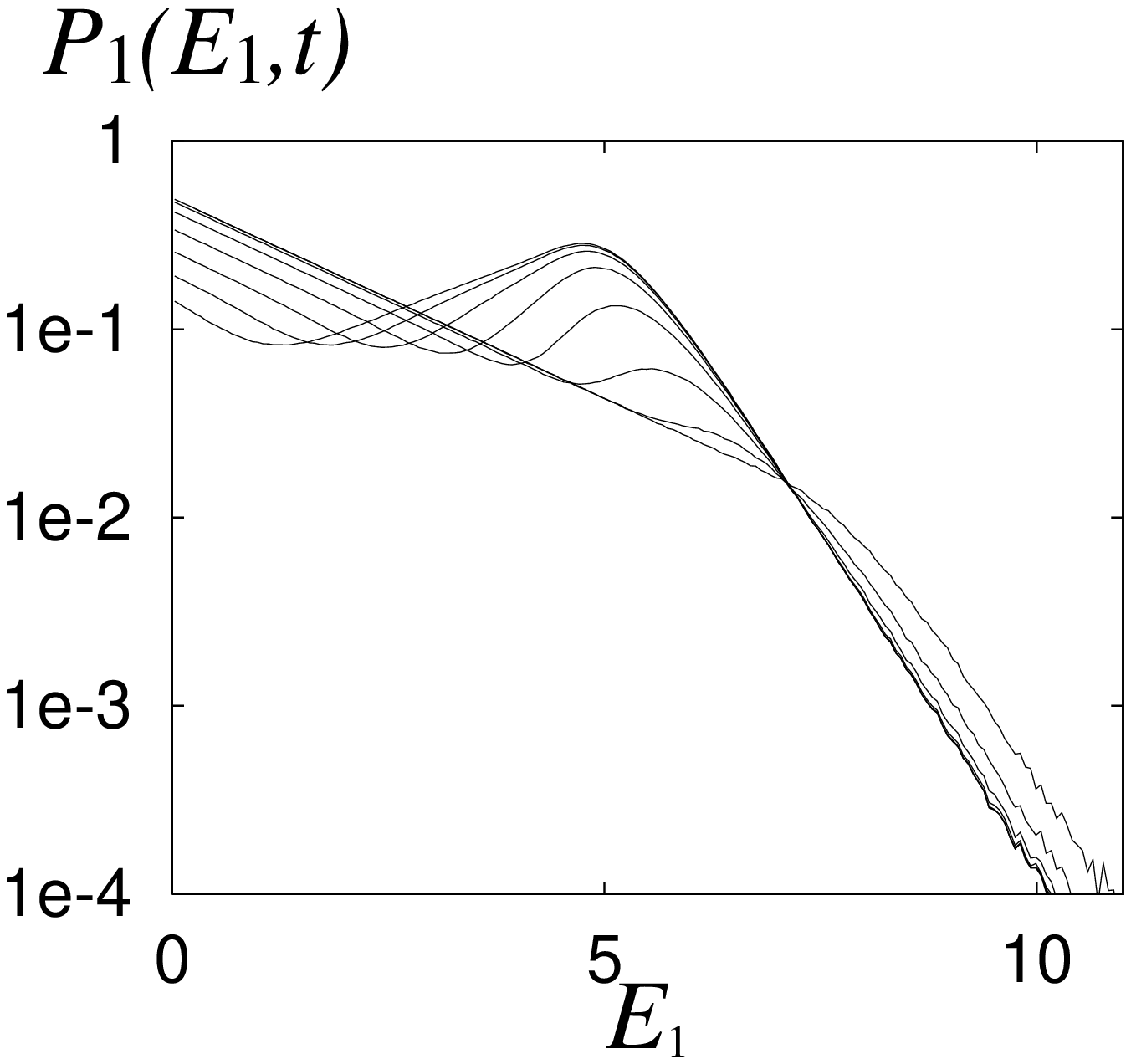,width=16.5cm}
\end{center}
\vspace{2cm}
\begin{center}
{\LARGE Fig.13}
\end{center}


\begin{thebibliography}{99}
\bibitem{ZFCexperiment1}P. Svedlindh, P. Granberg, P. Nordblad, L. Lundgren 
and H. S. Chen: Phys. Rev B {\bf 35}(1987)268.

\bibitem{ZFCexperiment2}L. Lundgren, P. Svedlindh and O. Beckman: 
Phys. Rev B {\bf 26}(1982)3990.

\bibitem{TRMexperiment}Ph. Refreggier, E. Vincent, J. Hammann and M. Ocio: 
J. Phys. (Paris) {\bf 48}(1987)1533.

\bibitem{ZFCTS}P. Granberg, L. Sandlund, P. Nordblad, P. Svedlindh and 
L. Lundgren: Phys. Rev. B {\bf 38}(1988)7097.

\bibitem{ZFCTC}P. Granberg, L. Lundgren and P. Nordblad: J. Magn. Magn. Mater 
{\bf 92}(1990)228.

\bibitem{MCS}J. O. Andersson, J. Mattsson and P. Svedlindh: Phys. Rev. B 
{\bf 46}(1992)8297.

\bibitem{acTC1}E. Vincent, J. P. Bouchaud, J. Hammann and F. Lefloch: 
Phil. Mag. B {\bf 71}(1995)489.

\bibitem{acTC2}F. Lefloch, J. Hammann, M. Ocio and E. Vincent: 
Europhys. Lett {\bf 18}(1992)647.

\bibitem{acTC3}J. O. Andersson, J. Mattsson and P. Nordblad: 
Phys. Rev. B {\bf 48}(1993)13977.

\bibitem{droplet1}D. S. Fisher and D. A. Huse: Phys. Rev. B {\bf 38}(1988)373.

\bibitem{droplet2}G. J. M. Koper and H. J. Hilhorst: J. Phys. (Paris) 
{\bf 49}(1988)429.

\bibitem{hierarchical1}M. Sasaki and K. Nemoto: J. Phys. Soc. Jpn 
{\bf 68}(1999)1148. 

\bibitem{hierarchical2}H. Yoshino: J. Phys. A {\bf 30}(1997)1143.

\bibitem{hierarchical3}P. Sibani and K. H. Hoffmann: Phys. Rev. Lett 
{\bf 63}(1989)2853.

\bibitem{hierarchical4}C. Shulze, K. H. Hoffmann and P. Sibani: 
Europhys. Lett {\bf 15}(1991)361.

\bibitem{Bouchaud}J. P. Bouchaud and D. S. Dean: 
J. Phys. I France {\bf 5}(1995)265.

\bibitem{MREM2}B. Derrida: J. Phys. Lett. France {\bf 46}(1985)401.

\bibitem{MREM3}B. Derrida and E. Gardner: J. Phys. C {\bf 19}(1986)2253.

\bibitem{Jonason}K. Jonason, E. Vincent, J. Hammann, J. P. Bouchaud and 
P. Nordblad: Phys. Rev. Lett. {\bf 81}(1998)3243.

\bibitem{Nemoto}K. Nemoto: J. Phys. A {\bf 21}(1988)L287.

\bibitem{DPinDP}J. Villain: J. Phys. France {\bf 46}(1985)1843.

\end{thebibliography}
\end{document}